\documentclass[journal,transmag]{IEEEtran}
\usepackage{amsmath}
\usepackage{graphicx,color}

\ifCLASSINFOpdf
\else
\fi
\usepackage{amsmath}
\begin{document}
%
\title{Quantifying Magnetic Fields Using Deformed Diamagnetic Liquid Profiles}


\author{\IEEEauthorblockN{David Shulman\IEEEauthorrefmark{1},\IEEEauthorrefmark{2}
}
\IEEEauthorblockA{\IEEEauthorrefmark{1}Department of Chemical Engineering, Ariel University, Ariel, Israel 407000}
\IEEEauthorblockA{\IEEEauthorrefmark{2}Physics Department, Ariel University, Ariel 40700, Israel}

\thanks{ 
Corresponding author: D.Shulman (email: davidshu@ariel.ac.il).}}

%



\IEEEtitleabstractindextext{%
\begin{abstract}
Measuring the magnetic field of permanent magnets can be challenging, but recent research has demonstrated the potential of using deformed diamagnetic liquids to estimate the magnetic field. In this paper, we explore two methods for measuring the magnetic field from the response of the diamagnetic liquid. The first method involves measuring the profile of the deformed liquid with a laser and then calculating the square of the magnetic field using an appropriate equation. The second method involves measuring the maximum slope of the liquid and numerically calculating the magnetic field distribution using the model of an ideal solenoid. We present experimental results using these methods and compare them with other established methods for measuring magnetic fields. The results show that the proposed methods are effective and have potential for use in a variety of applications. The proposed methods can help address the challenge of measuring magnetic fields in situations where other methods are not suitable or practical.
\end{abstract}

}

\maketitle

\IEEEdisplaynontitleabstractindextext

%
\IEEEpeerreviewmaketitle

\section{Introduction}

Magnetic field measurement is a fundamental task in various fields, such as materials science, engineering, and medicine. Traditional methods for magnetic field measurement include Hall probes \cite{hall_effect}, fluxgate magnetometers \cite{primdahl1979fluxgate}, and superconducting quantum interference devices (SQUIDs) \cite{clarke1989principles}. These methods are highly accurate and sensitive, but they can be expensive, require sophisticated instrumentation, and may not be suitable for non-destructive testing or in situ measurements.

In this paper a novel method for measuring magnetic fields has been proposed based on the response of a deformed diamagnetic liquid. Diamagnetic materials are those that exhibit a weak, negative response to magnetic fields and are repelled by the poles of a magnet. When a diamagnetic liquid is subjected to a magnetic field, it deforms into a characteristic shape that depends on the strength and direction of the field. By measuring the shape of the deformed liquid, it is possible to infer the distribution of the magnetic field that caused the deformation.

Various studies have explored the deformation of diamagnetic liquids under magnetic fields, and the accuracy and reliability of this method have been evaluated \cite{gendelman2019study,shulman2022measurement,shulman2022forces,chen2011deformation,bormashenko2019moses}. However, to the best of our knowledge, there has been no previous work on measuring the magnetic field from the response of a deformed diamagnetic liquid.

In this paper, we propose a method for measuring the magnetic field of a permanent magnet based on the response of a deformed diamagnetic liquid. We present the theoretical background of the method, experimental setup and procedures, and the results of our experiments. We also compare our results with those obtained from traditional magnetic field measurement methods and discuss the advantages and limitations of our method. Our method has the potential to be a low-cost, non-destructive, and non-invasive alternative for magnetic field measurement, particularly for large or irregularly-shaped magnets. 

The rest of the paper is organized as follows. In Section II, we provide an overview of the theoretical background of the method. In Section III, we describe the experimental setup and procedures. In Section IV, we present the results of our experiments. In Section V, we compare our results with those obtained from traditional magnetic field measurement methods. Finally, in Section VI, we discuss the advantages and limitations of our method and provide some concluding remarks.

\section{Theoretical Background of the Methods}
\subsection{Theoretical background of the first method}

The first method for measuring magnetic fields involves using the response of a deformed diamagnetic liquid to infer the distribution of the magnetic field that caused the deformation. Diamagnetic materials, such as certain types of liquids, exhibit a weak negative response to magnetic fields and are repelled by the poles of a magnet. When a diamagnetic liquid is subjected to a magnetic field, it deforms into a characteristic shape that depends on the strength and direction of the field.

\begin{equation}
z\left( r,h\right) =\frac{\chi B^{2}\left( r,h\right) }{2\mu _{0}\rho g},
\label{eqn:profile}
\end{equation}

and equation for the square of the magnetic field:
\begin{equation}
  B^{2}=z\left( r,h\right)\frac{2 \mu _{0}\rho g }{\chi},
\label{eqn:sqField}
\end{equation}

where $z$ is the height of the deformed liquid, $r$ is the radial distance from the center of the deformed liquid, $h$ is the separation between the magnet and liquid/vapor interface, $\chi$ is the magnetic susceptibility of the liquid, $B$ is the magnetic field strength, $\mu_0$ is the magnetic permeability of free space, $\rho$ is the density of the liquid, and $g$ is the acceleration due to gravity.

If the surface tension of the liquid can be ignored, the shape of the deformed liquid can be described by Equation (\ref{eqn:profile}). This equation relates the height of the liquid surface at any point to the strength of the magnetic field at that point. The square of the magnetic field can then be calculated using Equation (\ref{eqn:sqField}). These equations allow the magnetic field distribution to be inferred from the shape of the liquid surface, which can be measured with a laser, see Fig. (\ref{Setup_N}).

This method has the advantage of being non-invasive and contactless, and it can be used to measure the magnetic field distribution in a wide range of applications. However, if the surface tension of the liquid cannot be ignored, the equations become more complex, and the method may require additional corrections. Additionally, the accuracy of the method may be affected by factors such as the surface cleanliness of the liquid and the presence of other nearby magnetic objects.

\subsection{Theoretical Background of the Second Method}

The second technique for measuring the magnetic field involves examining the shape of a liquid surface that has been distorted by the magnetic field. The method consists of measuring the maximum slope of the distorted profile, and then comparing the experimental measurement to the analytical solution for the maximum slope using numerical methods.

In order to calculate the magnetic field using this method, a model of the magnetic field must be chosen e.g. from the Biot-Savart law for the ideal solenoid \cite{shulman2023semi}. The Biot-Savart law states that the magnetic field at a point in space, due to a current-carrying wire, is proportional to the current and the length of the wire, and inversely proportional to the distance from the wire. The Biot-Savart law for an ideal solenoid, which is a coil of wire wound in a helix with a uniform current density, can be derived by integrating the Biot-Savart law over the entire length of the solenoid.

Once the model of the magnetic field has been chosen, it can be compared to the experimental measurement of the maximum slope of the distorted liquid surface. By using numerical methods to fit the model to the experimental data, the magnetic field can be quantified more precisely.
Please see in the appendix the complete mathematical formulation of this method.
\section{Apparatus and Measurement Technique}
\subsection{Apparatus}
The experimental setup used to measure the magnetic field strength consists of the following components:
\begin{itemize}
  \item A sample liquid placed in a Petri dish with a diameter of 90 mm.
  \item A helium-neon laser (4mW 1107p) with a wavelength of 633 nm, supplied by JDS Uniphase Corporation, to enable the measurement of the shape of the liquid/vapor interface.
  \item Stacks of Neodymium permanent magnets, supplied by MAGSY, Czech.
\item An XYZ actuator with an accuracy of 10 $\mu m$ for precise location of the permanent magnet. The actuator was adjusted from components supplied by CCM Automation Technology. In the experiments, an Arduino controller for step motors was used.
  \item A digital camera (8.0-megapixel digital bridge camera Sony Cyber-shot DSC-F828).
  \item A Gauss meter, GM2 Gauss Meter, manufactured by AlphaLab Inc., USA, with an accuracy of ±0.01 T.
\end{itemize}

The experiments were carried out under ambient conditions (P=1 atm; T=25 C). A photograph of the experimental unit is shown in Fig. (\ref{Setup_photo}). The angle of incidence laser beam on the surface of the liquid is about 5°.
\subsection{Measurement Technique}
\subsubsection{Introduction}
The measurement technique used in this study is based on the observation of the shape of the liquid/vapor interface using a laser displacement sensor. The shape of the interface is related to the magnetic field strength in the region above the permanent magnet. As the magnet is moved closer to the liquid surface, the magnetic field strength increases, which leads to a change in the shape of the interface. This change in shape can be captured using the laser and analyzed to determine the magnetic field strength.

The experimental setup consists of a Petri dish filled with the liquid sample, which is placed on a stable surface. The Neodymium permanent magnet is then placed at various distances above the surface of the liquid, and the resulting shape of the liquid/vapor interface is observed using a helium neon laser with a wavelength of 633 nm. The laser beam is directed at the liquid surface at an angle of incidence of about 5 degrees, and the reflected beam on the screen is captured by a digital camera.

When the fluid is deformed by a magnetic field by an angle $\Delta \theta $,
the reflection angle shifts by $2\Delta \theta $, which causes a
corresponding shift in height $\Delta y$ on the screen, as shown in Fig. (\ref{Setup_N}).
Hence we have the following relationship for the change in the angle of
reflection:

\begin{equation}
2\Delta \theta =\Theta-\arctan \left( \frac{y-\Delta y}{L}\right)
\label{change}
\end{equation}

The angle of displacement is directly related to the slope of the liquid/vapor interface, which can be used to calculate the magnetic field strength using the Young-Laplace equation. The accuracy of the measurements is dependent on the accuracy of the laser displacement sensor, as well as the accuracy of the positioning of the magnet.

To ensure accurate measurements, the position of the magnet was fixed with a laboratory-built XYZ actuator with an accuracy of 10 micrometers. The accuracy of the magnet position was verified using a Gauss meter with an accuracy of ±0.01 T. The experiments were carried out under ambient conditions of temperature and pressure (T=25 C, P=1 atm).
\subsubsection{Measurement Technique for the First Method}
To calculate the surface profile from the reflection angle from a curved surface, we must obtain the displacement of the water as a function of position. The small angle approximation allows us to consider the change in the angle of the water surface as the slope of the water. This makes it possible to obtain the displacement by performing a Riemann sum on the measured data. In other words, we can write:

\begin{equation}
\Delta z(r)=\sum_{i=1}^n \left[\tan \left(2\Delta \theta(r_i)\right) \Delta r_i\right]
\end{equation}

where $\Delta z(r)$ is the displacement of the water surface at position $r$, $\Delta \theta(r_i)$ is the change in reflection angle at the $i$th measurement point, and $\Delta r_i$ is the distance between consecutive measurement points. The summation is taken over $n$ measurement points.

\subsubsection{Measurement Technique for the Second Method}

The measurement technique for the second method involves measuring the maximum slope of the liquid surface distorted by the magnetic field. This maximum slope can be obtained by analyzing the shift in the reflection angle on the screen caused by the curved surface.

To perform this measurement, the magnet is moved above the surface of the liquid, and the resulting shift in the reflection angle on the screen is recorded. The maximum shift in the reflection angle corresponds to the point where the slope of the liquid surface is at its maximum. This method provides a more simple measurement of the magnetic field compared to the first method, as it directly measures the maximum slope of the liquid surface.

Once the maximum shift in the reflection angle is recorded, it can be used to calculate the maximum slope of the liquid surface using Eq. (\ref{change}).

\section{Results and Discussion}
\subsection{First Method}
\begin{figure*}[!t]
\normalsize
  \begin{minipage}{0.48\textwidth}
     \centering
     \includegraphics[width=1\linewidth]{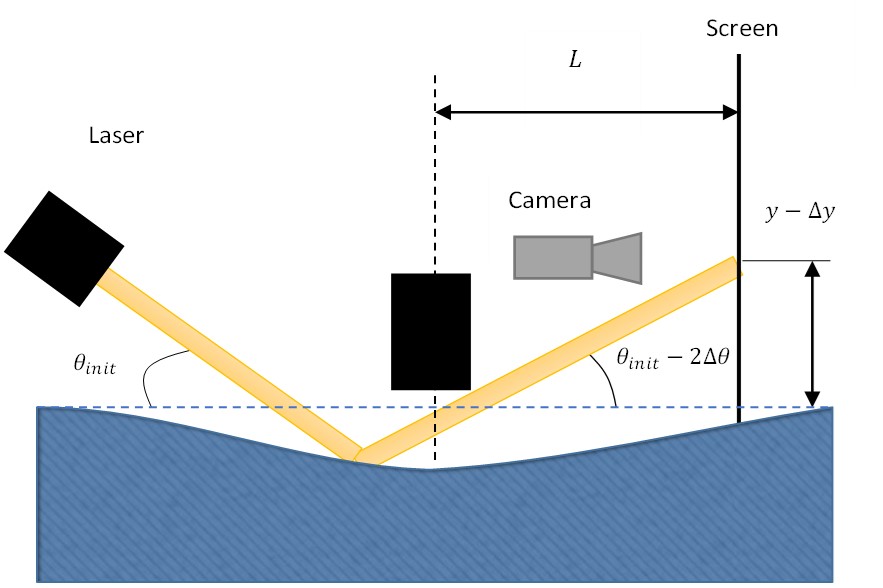}
     \caption{The schematic outline of the experiment setup}\label{Setup_N}
  \end{minipage}\hfill
  \begin{minipage}{0.48\textwidth}
     \centering
     \includegraphics[width=1.2\linewidth]{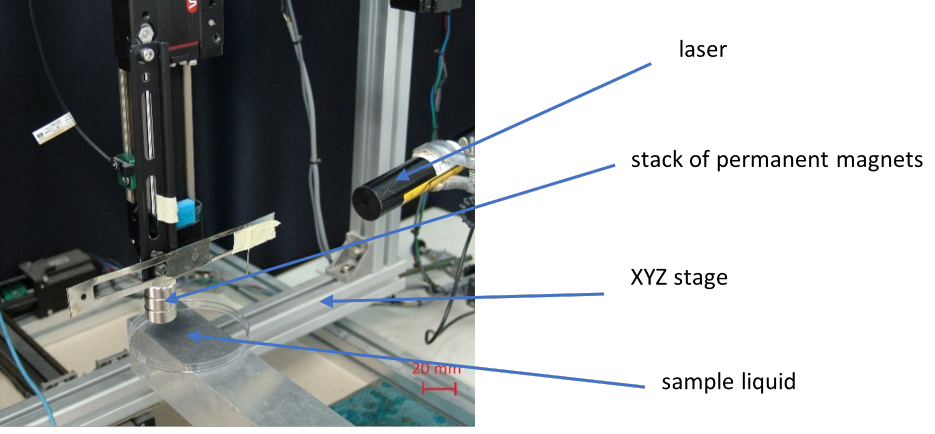}
     \caption{The photograph of the experimental setup}\label{Setup_photo}
  \end{minipage}
\hrulefill
\vspace*{4pt}  
  \begin{minipage}{0.48\textwidth}
     \centering
\includegraphics[width=\linewidth]{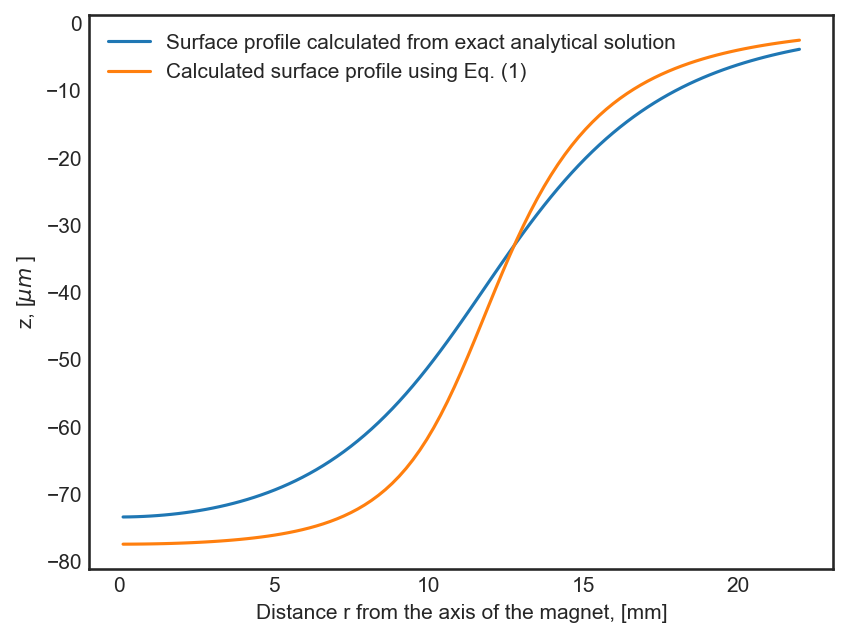}
\caption{Comparison of the surface profile calculated using Eq. (\ref{eqn:profile}) and the exact analytical solution Eq. (\ref{z}) on the case of water.}
\label{fig:1}
  \end{minipage}\hfill
  \begin{minipage}{0.48\textwidth}
     \centering
\includegraphics[width=\linewidth]{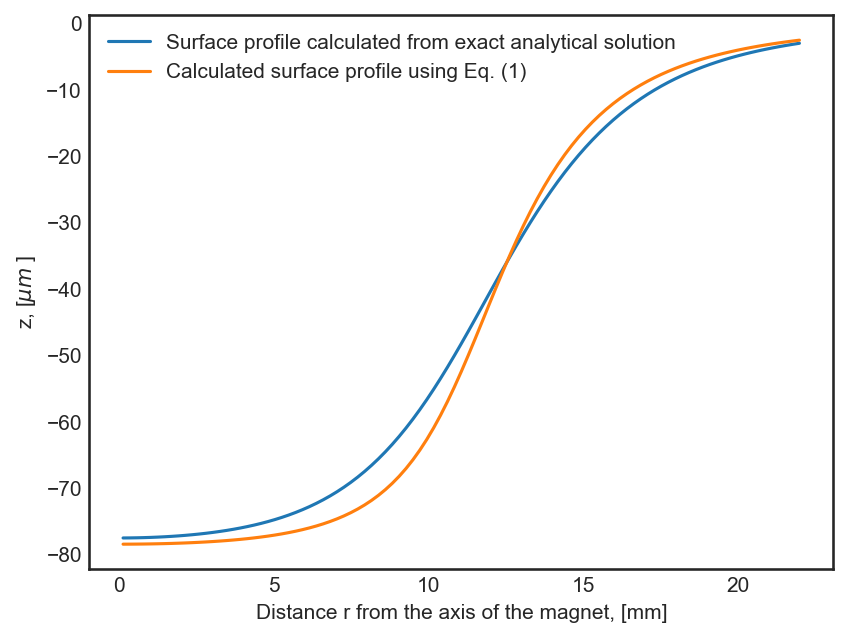}
\caption{Comparison of the surface profile calculated using Eq. (\ref{eqn:profile}) and the exact analytical solution Eq. (\ref{z}) on the case of ethanol.}
\label{fig:2}
  \end{minipage}
\vspace*{4pt} 
  \begin{minipage}{0.48\textwidth}
     \centering
\includegraphics[width=\linewidth]{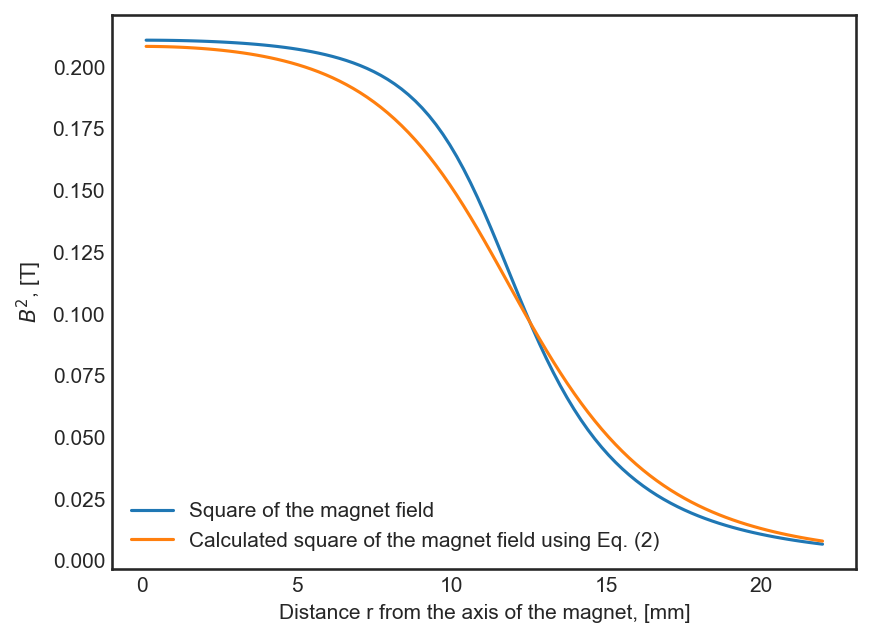}
\caption{Comparison of the calculated and exact square of magnetic field for ethanol, using Eq. (\ref{eqn:sqField})}
\label{fig:3}
  \end{minipage}\hfill
  \begin{minipage}{0.48\textwidth}
     \centering
\includegraphics[width=\linewidth]{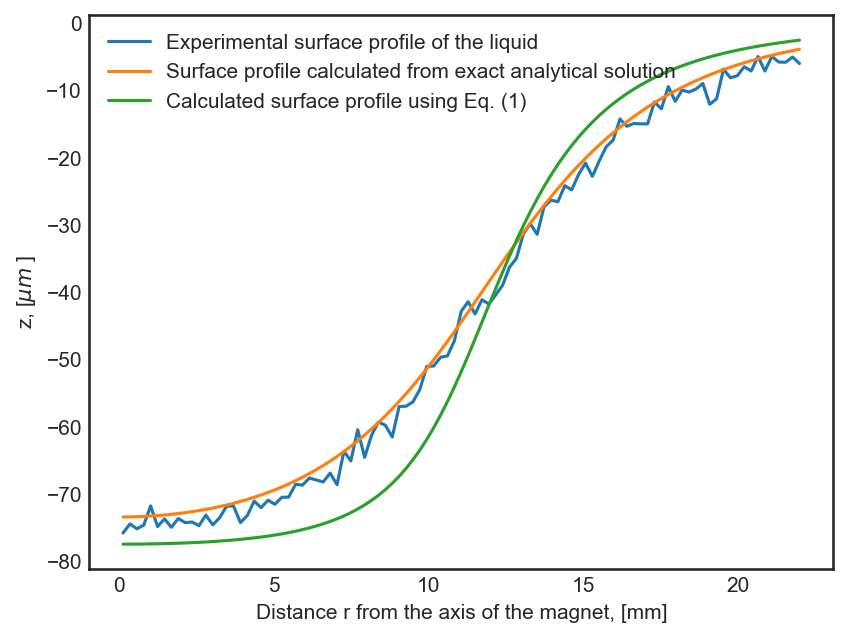}
\caption{Comparison of the surface profile calculated using Eq. (\ref{eqn:profile}) and the experimental surface profile obtained using a laser on the case of water. The distance between the magnet and the liquid surface is 3 mm. }
\label{fig:4}
  \end{minipage}
\vspace*{4pt} 
\end{figure*}

Fig. (\ref{fig:1}) compares the surface profile of the diamagnetic liquid, calculated using Eq. (\ref{eqn:profile}), to the exact analytical solution given by Eq. (\ref{z}), in the case of water. The magnet was positioned 3 mm above the liquid surface during the measurement. The distance between the magnet and the liquid surface is 3 mm.
As can be seen from the figure, there is a good agreement between the two solutions. However, it should be noted that the comparison could be even better with a liquid having a smaller surface tension. For example, when using ethanol or water with added surfactants such as dish soap or detergent, the comparison is expected to be even more accurate, as demonstrated in Fig. (\ref{fig:2}), which shows the comparison of the surface profile of ethanol obtained using this method.
In Fig. (\ref{fig:3}), we compare the exact square of the magnetic field in ethanol to the square of the magnetic field calculated using Eq. (\ref{eqn:sqField}). 

The comparison of the calculated and exact solutions for the square of the magnetic field in the case of ethanol shows a high degree of agreement, with an R-squared value of 0.99 indicating very good correspondence. However, it should be noted that there is a maximum error of approximately $9\%$ for the $B^2$ values, which should be taken into consideration for specific applications.

Fig. (\ref{fig:4}) depicts a comparison of the surface profile of water obtained from experimental data and the profile calculated using Eq. (\ref{eqn:profile}). As seen from the figure, there is some noise in the experimental data, but the calculated profile from the analytical solution is very close to the experimental data. Based on this observation, we can conclude that the main source of error in this method is due to the neglect of surface tension, which is a critical factor in determining the shape of the liquid surface. Therefore, future research in this area should focus on developing a more accurate model that accounts for the effect of surface tension, which would improve the accuracy of the method and extend its applicability to a wider range of liquids. Nevertheless, the current results show that the proposed method is effective for measuring the magnetic field and has potential for use in various applications where other methods may not be suitable or practical.

\subsection{Second Method}
\begin{figure*}[!t]
\normalsize
  \begin{minipage}{0.48\textwidth}
     \centering
     \includegraphics[width=\linewidth]{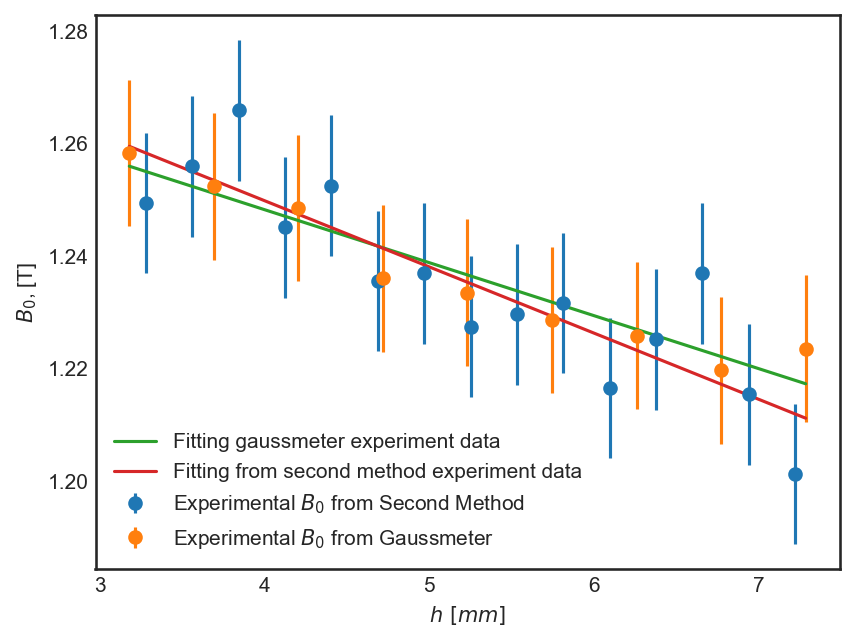}
     \caption{Comparison of the coefficient $B_0$ obtained from the second method and using the Gaussmeter for a permanent magnet. }\label{fig:5}
  \end{minipage}\hfill
  \begin{minipage}{0.48\textwidth}
     \centering
     \includegraphics[width=\linewidth]{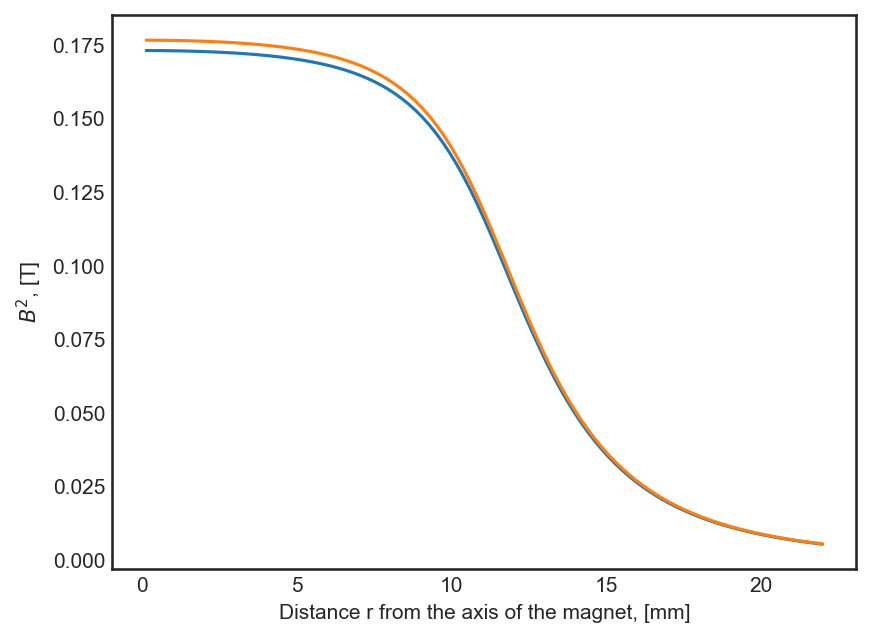}
     \caption{Comparison of the square of the magnetic field distribution for an exact value of $B_0$ and a deviation of 1\%. The plot shows a visualization of the difference between the two distributions, with a maximum error of 2\% in the case of deviation.}\label{fig:6}
  \end{minipage}

\end{figure*}

The second method for measuring the magnetic field involved measuring the maximum slope of the curved surface, followed by fitting the model of the magnetic field to the data. To verify the accuracy of this method, we also measured the distribution of the magnetic field using a gaussmeter (see Ref. \cite{shulman2023semi}) and compared the results to those obtained from this method, see Fig. (\ref{fig:5}).

In our framework, we use a model for the magnetic field derived from the Biot-Savart law, and the necessary parameters for this model are the magnetization $B_0$ that is obtained experimentally. In Fig. (\ref{fig:5}), we show the experimentally obtained values for $B_0$ as a function of vertical distance from the magnet. The error bars provide an estimate of the average measurement error associated with the instrument used. It can be observed from the figure that there are several outlier points that deviate significantly from the overall trend. Therefore, to obtain more accurate results, it is necessary to take the mean of multiple measurements and perform statistical analysis to identify and remove outliers. The plot shows that the magnetization $B_0$ changes as a function of distance from the magnet. This is consistent with the presence of inhomogeneity, as previously discussed in Ref. \cite{shulman2023semi}.
The experimental data in Fig. (\ref{fig:5}) is fitted with a linear function, which is also shown in the figure. The good comparison between the fitted function and the experimental data confirms the accuracy of the measurements. 

Upon comparing the experimental values of the magnetic field obtained from the second method with those obtained from the Gaussmeter, we observed that the maximum difference between the two sets of values was less than 1\%. This result indicates that the second method is a reliable and accurate way of measuring the magnetic field of permanent magnets. Furthermore, the error bars in the experimental data represent the average error of the measuring apparatus, which is relatively small. However, we also observed that the magnetization $B_0$ obtained experimentally varies with the distance h from the magnet, as described in Ref. \cite{shulman2023semi}.
Fig. (\ref{fig:6}) provides a visualization of the difference in the distribution of the square of the magnetic field in the case of a deviation of $B_0$ of 1\%. As can be seen from the figure, the maximum error in $B^2$ in this case is about 2\%. These results indicate that the method is relatively robust to small variations in $B_0$, which is an important consideration in practical applications where accurate measurement of the magnetic field is critical.
\subsection{Pros and Cons of Both Methods}
There are two methods for measuring the magnetic field using deformed diamagnetic liquids: the first method involves measuring the profile of the deformed liquid, while the second method involves measuring the maximum slope of the liquid. Each method has its own advantages and disadvantages.

\subsubsection{Method 1}
Pros:
\begin{enumerate}
\item The mathematical calculation is simple.
\item The method does not require knowledge of the surface tension of the liquid used for measurement.
\end{enumerate}

Cons:
\begin{enumerate}
\item The measurement technique can be difficult.
\item The method may be less accurate compared to the second method.
\end{enumerate}

\subsubsection{Method 2}
Pros:
\begin{enumerate}
\item The measurement technique is simple.
\item The method is very accurate.
\end{enumerate}

Cons:
\begin{enumerate}
\item The mathematical calculation can be difficult.
\item The method requires knowledge of the surface tension of the liquid used for measurement.
\end{enumerate}

By considering the pros and cons of both methods, it is possible to choose the appropriate method based on the requirements of the specific application. For example, if accuracy is the primary concern and the surface tension of the liquid is known, then the second method may be preferred. Conversely, if simplicity is more important and accuracy is less of a concern, then the first method may be more suitable.
\section{Conclusion}

In this study, we explored two methods for measuring the magnetic field of permanent magnets using deformed diamagnetic liquids. The first method involved measuring the profile of the deformed liquid with a laser and calculating the square of the magnetic field using an appropriate equation. The second method involved measuring the maximum slope of the liquid and numerically calculating the magnetic field distribution using the model of an ideal solenoid.

We presented experimental results using these methods and compared them with other established methods for measuring magnetic fields. The results showed that both methods were effective and had potential for use in a variety of applications.

The first method, despite its simplicity in mathematical calculation, requires a difficult measurement technique and is less accurate compared to the second method. On the other hand, the second method, while requiring more difficult mathematical calculations, provides very accurate results and has a simpler measurement technique. However, it requires knowledge of the surface tension of the measuring liquid.

Overall, these methods can help address the challenge of measuring magnetic fields in situations where other methods are not suitable or practical. Future work could focus on exploring other possible applications and further improving the accuracy of these methods.
\onecolumn  
\appendix
\section*{Theoretical background of the second method}

The Young-Laplace equation, which describes the equilibrium shape of a liquid surface in response to external forces, including magnetic forces. The left-hand side of the equation represents the gravitational and surface tension forces, while the right-hand side represents the magnetic force, see Ref. \cite{gendelman2019study}:

\begin{equation}
\frac{\partial ^{2}z}{\partial r^{2}}+\frac{1}{r}\frac{\partial z}{\partial r%
}-\frac{\rho g}{\gamma }z=\frac{\chi B^{2}\left( r,h\right) }{2\mu
_{0}\gamma }  \label{z_equation}
\end{equation}
The solution to this equation gives the profile of the deformed liquid surface: 
\begin{equation}
z\left( r,h\right) =-\left[ \int_{0}^{r}\frac{\chi B^{2}\left( r^\prime,h\right) }{%
2\mu _{0}\gamma }I_{0}\left( \lambda _{c}^{-1}r^\prime\right) r^\prime dr^\prime\right]
K_{0}\left( \lambda _{c}^{-1}r\right)-\left[ \int_{r}^{\infty }\frac{\chi
B^{2}\left( r^\prime,h\right) }{2\mu _{0}\gamma }K_{0}\left( \lambda
_{c}^{-1}r^\prime\right) r^\prime dr^\prime\right] I_{0}\left( \lambda _{c}^{-1}r\right) \label{z}
\end{equation}

where $\gamma$ is the surface tension of the liquid, $I_0$ and $K_0$ are modified Bessel functions of the first and second kind, respectively. The interplay between the gravity and the surface tension is quantified by the capillary length, denoted $\lambda$. The derivative of Eq. \ref{z} is given by:
\begin{equation}
\theta \approx \frac{dz}{dr}=\left[ \int_{0}^{r}\frac{\chi B^{2}\left( r^\prime,h\right) }{%
2\mu _{0}\gamma }I_{0}\left( \lambda _{c}^{-1}r^\prime\right) r^\prime dr^\prime\right] \lambda
_{c}^{-1}K_{1}\left( \lambda _{c}^{-1}r\right) -\left[ \int_{r}^{\infty }%
\frac{\chi B^{2}\left( r^\prime,h\right) }{2\mu _{0}\gamma }K_{0}\left( \lambda
_{c}^{-1}r^\prime\right) r^\prime dr^\prime\right] \lambda _{c}^{-1}I_{1}\left( \lambda
_{c}^{-1}r\right)  \label{theta}
\end{equation}
To find the maximum slope of the deformed liquid surface, we apply the second derivative to the equation for the curved surface (Eq. \ref{theta}), set it equal to zero, and then solve numerically and fit to the model of the magnetic field derived from the Biot-Savart law for the ideal solenoid. This allows us to calculate the magnetic field strength at any point in the solenoid, and is the basis for the second method of measuring the magnetic field.
\begin{multline}
\frac{d^{2}z}{dr^{2}} =\frac{\chi B^{2}\left( r,h\right) }{2\mu _{0}\gamma 
}-\left[ \int_{0}^{r}\frac{\chi B^{2}\left( r,h\right) }{2\mu _{0}\gamma }
I_{0}\left( \lambda _{c}^{-1}r\right) rdr\right] \lambda _{c}^{-2}\left(
K_{0}\left( \lambda _{c}^{-1}r\right) +\frac{1}{\lambda _{c}^{-1}r}
K_{1}\left( \lambda _{c}^{-1}r\right) \right) -\\ 
-\left[ \int_{r}^{\infty }\frac{\chi B^{2}\left( r,h\right) }{2\mu
_{0}\gamma }K_{0}\left( \lambda _{c}^{-1}r\right) rdr\right] \lambda
_{c}^{-2}\left( I_{0}\left( \lambda _{c}^{-1}r\right) -\frac{1}{\lambda
_{c}^{-1}r}I_{1}\left( \lambda _{c}^{-1}r\right) \right)  \notag
\end{multline}

and hence
\begin{multline}
B^{2}\left( r_{m},h\right)  = \left[ \int_{0}^{r_{m}}B^{2}\left( r,h\right)
I_{0}\left( \lambda _{c}^{-1}r\right) rdr\right] \lambda _{c}^{-2}\left(
K_{0}\left( \lambda _{c}^{-1}r_{m}\right) +\frac{1}{\lambda _{c}^{-1}r_{m}}%
K_{1}\left( \lambda _{c}^{-1}r_{m}\right) \right) -\\
-\left[ \int_{r_{m}}^{\infty }B^{2}\left( r,h\right) K_{0}\left( \lambda
_{c}^{-1}r\right) rdr\right] \lambda _{c}^{-2}\left( I_{0}\left( \lambda
_{c}^{-1}r_{m}\right) -\frac{1}{\lambda _{c}^{-1}r_{m}}I_{1}\left( \lambda
_{c}^{-1}r_{m}\right) \right)  \notag \label{ddz}
\end{multline}

Where $r_m$ is a radial distance from magnet z axes to a point of the maximum slope curved liquid surface.

The model of the magnetic field used in this investigation is the solution of the Biot-Savart law for the ideal solenoid:

\begin{equation}
B_{r}\left( r,h\right) =B_{0}\int_{0}^{\pi /2}d\psi \left( \cos ^{2}\psi
-\sin ^{2}\psi \right) \left\{ \frac{\alpha _{+}}{\sqrt{\cos ^{2}\psi
+k_{+}^{2}\sin ^{2}\psi }}-\frac{\alpha _{-}}{\sqrt{\cos ^{2}\psi
+k_{-}^{2}\sin ^{2}\psi }}\right\}
\end{equation}

\begin{equation}
B_{z}\left( r,h\right) =\frac{B_{0}a}{r+a}\int_{0}^{\pi /2}d\psi \left( 
\frac{\cos ^{2}\psi +\tau \sin ^{2}\psi }{\cos ^{2}\psi +\tau ^{2}\sin
^{2}\psi }\right) \left\{ \frac{\beta _{+}}{\sqrt{\cos ^{2}\psi
+k_{+}^{2}\sin ^{2}\psi }}-\frac{\beta _{-}}{\sqrt{\cos ^{2}\psi
+k_{-}^{2}\sin ^{2}\psi }}\right\}
\end{equation}

\begin{equation*}
\alpha _{\pm }=\frac{a}{\sqrt{h_{\pm }^{2}+\left( r+a\right) ^{2}}},\ \ \ \
\ \ \ \ \ \beta _{\pm }=\frac{h_{\pm }}{\sqrt{h_{\pm }^{2}+\left( r+a\right)
^{2}}}\ \ \ \ \ \ 
\end{equation*}

\begin{equation*}
h_{+}=h,\ \ \ h_{-}=h-2b,\ \ \ \ \tau =\frac{a-r}{a+r}
\end{equation*}

\begin{equation*}
k_{\pm }=\sqrt{\frac{h_{\pm }^{2}+\left( a-r\right) ^{2}}{h_{\pm
}^{2}+\left( a-r\right) ^{2}}}
\end{equation*}

where $a$ is the radius and $2b$ is the length of the solenoid; $\left(r,\ \varphi,\ h\right)$ are the cylindrical coordinates with the origin at the center of the solenoid; $n$ – is the number of turns per unit length. To obtain the equations in the current form, we have also introduced the following integration variable change: $2\psi\equiv\pi-\varphi$. To compare the calculation results with the results of the measurements, the radius and the length of the solenoid for the calculations were chosen equal to the radius and the length of the permanent magnet investigated experimentally. 

Our problem involves two unknowns: the magnetization of the permanent magnet, $B_0$, and the radial distance of the maximum slope from the magnet axis, $r_m$. We have two equations that describe the problem:

\begin{equation}
    \begin{cases}
     \theta \approx \frac{dz}{dr}=\left[ \int_{0}^{r_{m}}\frac{\chi B^{2}\left( r^\prime,h\right) }{%
2\mu _{0}\gamma }I_{0}\left( \lambda _{c}^{-1}r^\prime\right) r^\prime dr^\prime\right] \lambda
_{c}^{-1}K_{1}\left( \lambda _{c}^{-1}r_{m}\right) -\left[ \int_{r_{m}}^{\infty }%
\frac{\chi B^{2}\left( r^\prime,h\right) }{2\mu _{0}\gamma }K_{0}\left( \lambda
_{c}^{-1}r^\prime\right) r^\prime dr^\prime\right] \lambda _{c}^{-1}I_{1}\left( \lambda
_{c}^{-1}r_{m}\right)\\ \\
    B^{2}\left( r_{m},h\right)  = \left[ \int_{0}^{r_{m}}B^{2}\left( r,h\right)
I_{0}\left( \lambda _{c}^{-1}r\right) rdr\right] \lambda _{c}^{-2}\left(
K_{0}\left( \lambda _{c}^{-1}r_{m}\right) +\frac{1}{\lambda _{c}^{-1}r_{m}}%
K_{1}\left( \lambda _{c}^{-1}r_{m}\right) \right) -\\
-\left[ \int_{r_{m}}^{\infty }B^{2}\left( r,h\right) K_{0}\left( \lambda
_{c}^{-1}r\right) rdr\right] \lambda _{c}^{-2}\left( I_{0}\left( \lambda
_{c}^{-1}r_{m}\right) -\frac{1}{\lambda _{c}^{-1}r_{m}}I_{1}\left( \lambda
_{c}^{-1}r_{m}\right) \right)  \notag
    \end{cases}   
\end{equation}
To solve this problem numerically, we use Brent's method \cite{brent2013algorithms}, which involves substituting the experimentally known value of $\theta$ into the equations and finding the zero of the resulting series. While we use Brent's method in our work, other suitable numerical methods can also be used.
\section*{Acknowledgment}

I would like to express my deep gratitude to Professor Meir Lewkowicz and Professor Edward Bormashenko, my research supervisors, for their patient guidance, enthusiastic encouragement, and useful critiques
\section*{AUTHOR DECLARATIONS}
\subsection*{Conflict of Interest}
The author has no conflicts to disclose.

\bibliographystyle{IEEEtran}
\bibliography{mybibfile}

\twocolumn
\newpage

\end{document}